\documentclass[aps,prl, 
twocolumn,
groupedaddress]{revtex4-2}

\usepackage{dsfont}
\usepackage{graphicx}
\usepackage{dcolumn}
\usepackage{braket}
\usepackage{hyperref}
\usepackage{amsmath}
\usepackage{amssymb}
\usepackage{tikz}

\begin{document}

\title{\textbf{Non-magnetic floating phases in frustrated Haldane chains with a single-ion anisotropy} 
}%

\author{Bowy M. La Rivi\`ere}%
\email{Contact author: b.m.lariviere@tudelft.nl}
\affiliation{ Kavli Institute of Nanoscience, Delft University of Technology, Lorentzweg 1, 2628CJ Delft, The Netherlands }
\author{Natalia Chepiga}%
\affiliation{  Rudolf Peierls Centre for Theoretical Physics, University of Oxford, Clarendon Laboratory, Oxford OX1 3PU, United Kingdom}

\date{\today}

\begin{abstract}
We investigate the effect of a single-ion anisotropy in the bilinear–biquadratic spin-1 $J_1$–$J_2$ chain, focusing on the quantum phase transitions out of the trimerized phase. Using large-scale density matrix renormalization group simulations, we uncover a rich phase diagram comprising five gapped phases and, remarkably, two critical floating phases. These incommensurate Luttinger liquid phases emerge from the proliferation of non-magnetic domains -- $S_i^z=0$ states and dimers -- within a trimerized background and are confined to the zero-magnetization sector, while magnetic excitations remain gapped. We show that transitions between the symmetry-protected topological Haldane phase and the floating phases are governed by a composite critical line with central charge $c=2$, consistent with the coexistence of magnetic Gaussian and non-magnetic critical modes. These results shed new light on the long-standing problem of the Haldane–trimerized transition.
\end{abstract}

\maketitle

Study of quantum critical phenomena in low-dimensional quantum magnets remains a long-term challenge \cite{ThierryGiamargi}. Competing interactions induce frustration and lead to new phases and quantum phase transitions. The prominent example is spin-1/2 $J_1-J_2$ Heisenberg chains with nearest-neighbor (NN) and next-nearest-neighbor (NNN)  interactions.  The chain is critical up to the point where NNN-coupling exceeds $J_2/J_1 \approx 0.2411$ \cite{okamoto_fluid-dimer_1992} beyond which the spontaneously-dimerized phase emerges,  including the exact Majumdar-Ghosh point \cite{majumdarNextNearestNeighborInteractionLinear1969}.
For half-integer spin chains with $S>1/2$ the phase diagram is much richer and the commensurate critical phase at small $J_2/J_1$ and fully-dimerized phase at large NNN-coupling are now separated by the partially dimerized gapped phase and more exotic, yet experimentally realizable, critical floating phase \cite{chepigaFloatingCriticalDimerized2020a,PhysRevB.105.174402,pddz-42x9,PhysRevB.109.155130,PhysRevB.104.184402}.
Both critical phases form Luttinger liquids (LL) with quasi-long-range order being either commensurate and incommensurate respectively.
In the floating phase the dominant incommensurate wave-vector is not fixed to a certain value, but changes continuously -- {\it floats} \cite{bakCommensuratePhasesIncommensurate1982a}. 
In half-integer spin chains both, commensurate and incommensurate, critical phases are magnetic with gapless excitations in all magnetization sectors \cite{chepigaFloatingCriticalDimerized2020a}. 

The physics of integer spin chains is drastically different \cite{affleckCriticalTheoryQuantum1987b}.
A prominent example is a spin-1 Haldane chain characterized by a finite bulk gap \cite{haldaneContinuumDynamics1D1983,affleck_rigorous_1987} and topologically protected  spin-1/2 edge states \cite{TKennedy_1990,PhysRevLett.65.3181,tasakiGroundState$S1$2025}.
The transition between the Haldane and dimerized phase, if continuous, is described by the Wess-Zumino-Witten (WZW) SU(2)$_2$ critical theory \cite{affleckCriticalTheoryQuantum1987b}.
The next-nearest-neighbor interaction, however, does not lead to the dimerized phase but stabilizes the NNN-Haldane phase -- a topologically trivial uniform phase with gapped excitations that can be understood as a pair of intertwined Haldane chains \cite{kolezhukConnectivityTransitionFrustrated2002,kolezhukVariationalDensitymatrixRenormalizationgroup1997}. The presence of this trivial phase next to the spontaneously dimerized one gives rise to the non-magnetic Ising transition \cite{chepigaDimerizationTransitionsSpin12016a,chepigaCommentFrustrationMulticriticality2016a,Chepiga_spin_3_chain}.
In contrast to WZW criticality, this transition occurs entirely within the singlet sector, with magnetic excitations remaining gapped across the transition \cite{chepigaDimerizationTransitionsSpin12016a}.
It is driven by rearrangements of spin dimers without breaking them into spinons.
In fact, Ising criticality is intrinsically included in the WZW  SU(2)$_2$ theory via conformal embedding,  also called a coset construction \cite{osti_5929972}. In the presence of single-ion anisotropy, $D (S_i^z)^2$, which reduces the symmetry of the Heisenberg model from SU(2) to U(1), lifting the embedding condition, the WZW SU(2)$_2$ critical point splits into separate Gaussian and Ising  transitions \cite{s53y-qmr4}.

Ising transition is a textbook example describing melting of the period-two phase, while transitions out of the period-three and period-four phases exhibit richer variety of critical phenomena, including three-state Potts and Ashkin–Teller points, as well as non-conformal chiral transitions and floating phases \cite{fendleyCompetingDensitywaveOrders2004a,chepiga_kibble-zurek_2021,maceiraConformalChiralPhase2022,2019arXiv190802068R,Chepiga_Hard_boson}. 
Many of these critical phenomena have been predicted and observed in experiments on one-dimensional Rydberg arrays \cite{bernienProbingManybodyDynamics2017,zhangProbingQuantumFloating2025b}. 
This raises the question whether similar critical phenomena arise in the largely unexplored regime of $n$-merization transitions in antiferromagnetic spin chains.


To answer this question we investigate the bilinear-biquadratic (BLBQ) spin-1 zig-zag ladder. The model was motivated by experiments on  NiGa$_2$S$_4$ \cite{ni} and has been reported to host an extended trimerized phase \cite{Corboz_trimerization}. The nature of the transitions into this phase is a long-standing open question \cite{lecheminantSUNSelfdualSine2006,Corboz_trimerization}. We look at this model in the presence of single-ion anisotropy $D$:
\begin{equation}\label{eq:Hamiltonian}
    \begin{aligned}
        \mathcal{H} = \sum_i 
        & J_1 \left[ \cos\theta\mathbf{S}_{i}\cdot \mathbf{S}_{i+1} + \sin\theta \left(\mathbf{S}_{i}\cdot \mathbf{S}_{i+1} \right)^2 \right]\\
        + & J_2 \left[ \cos\theta\mathbf{S}_{i}\cdot \mathbf{S}_{i+2} + \sin\theta \left(\mathbf{S}_{i}\cdot \mathbf{S}_{i+2} \right)^2 \right]
        +  D \left( S^z_i \right)^2 ,
    \end{aligned}
\end{equation}
where $J_{1,2}$ denote respectively NN and NNN bilinear-biquadratic interactions parameterized by a single parameter $\theta$. Without loss of generality we fix $J_1=1$ to set the energy scale. 
When $J_2=D=0$, the model is reduced to BLBQ chain with the Haldane phase stretching between the exactly solvable points: Takhtajan-Babudjian WZW SU(2)$_2$  at $\theta=-\pi/4$ \cite{takhtajan_1982,babujian_1982} and  Uimin-Lai-Sutherland (ULS) WZW SU(3)$_1$ at $\theta=\pi/4$ \cite{Uimin_SU3_model,Lai_SU3_model,Sutherland_SU3_model} that marks the beginning of the critical $c=2$ phase \cite{fathPeriodTriplingBilinearbiquadratic1991,xianSpontaneousTrimerizationSpin11993,reedGaplessExcitationsSpin11994,bursillDensityMatrixRenormalization1995,itoiExtendedMasslessPhase1997,schmittStaticDynamicStructure1998,lauchliSpinNematicsCorrelations2006a}. The  $J_2$ interaction preserves SU(3) symmetry along $\theta=\pi/4$, but ultimately destabilizes the $\mathrm{SU}(3)_1$ criticality as the chain undergoes spontaneous trimerization \cite{lecheminantSUNSelfdualSine2006,Corboz_trimerization}. To remain generic, we fix $\theta=0.2\pi$, away from the SU(3)-symmetric line.

For $J_2=0$ the effect of the single-ion anisotropy $D$ is well established. Strong $|D|$ eventually destroys the Haldane phase: for $D>0$, the system enters a trivial large-$D$ phase, dominated, up to quantum fluctuations, by $S_i^z=0$ states,  while for $D<0$, the $S_i^z=0$ component is effectively suppressed, reducing the system to an Ising antiferromagnet (AFM) \cite{PhysRevB.84.054451,PhysRevB.28.3914,PhysRevB.67.104401,PhysRevB.34.6372,PhysRevB.104.024409}.
The Haldane phase is separated from the large-$D$ and Ising-AFM phases by the topological Gaussian \cite{ercolessi,PhysRevB.67.104401,PhysRevB.84.220402,haldaneSpontaneousDimerization,nomuraCriticalProperties11994} and Ising transitions \cite{PhysRevB.34.6372,PhysRevB.67.104401,albuquerqueQuantumPhaseDiagram2009} correspondingly. In this work, we investigate how single-ion anisotropy destroys Haldane and trimerized phases for $J_2>0$. We address this problem using state-of-the-art density matrix renormalization group (DMRG) simulations \cite{whiteDensityMatrixFormulation1992a, whiteDensitymatrixAlgorithmsQuantum1993a,schollwoeckDensitymatrixRenormalizationGroup2011a} (see End Matter for details).

\begin{figure}[!t]
    \centering
    \includegraphics[width=0.5\textwidth]{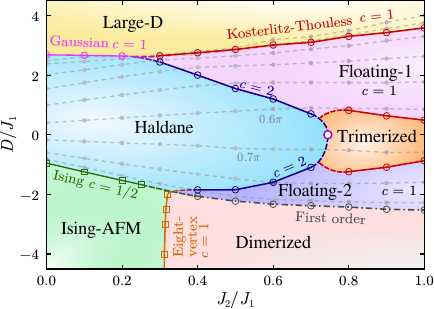}
    \caption{
        Phase diagram of the bilinear–biquadratic $J_1$–$J_2$ spin-1 chain with single-ion anisotropy $D$ [Eq.~\eqref{eq:Hamiltonian}] at $\theta=0.2\pi$. It contains five gapped phases: Haldane, large-$D$, and three symmetry-broken phases — trimerized, dimerized, and Ising antiferromagnetic (AFM). In addition, two floating phases appear: critical Luttinger liquids with central charge $c=1$, gapped magnetic excitations and incommensurate quasi-long-range order.
The trimerized phase is separated from both floating phases by Kosterlitz–Thouless (KT) transitions (red lines). Another KT transition separates Floating-1 from the large-$D$ phase. Dashed gray lines mark constant wave-vector $q$ (step $0.1\pi$); along $D=0$, correlations are commensurate with $q\approx 2\pi/3$ within numerical accuracy.
 Transitions between the Haldane and floating phases (blue lines) are characterized by $c=2$ and interpreted as a coexistence of two liquids magnetic and non-magnetic. The eight-vertex (Gaussian) transition separates the Ising-AFM and dimerized phases.  
Along D=0, our results are consistent with a direct continuous transition(dashed blue line)  with central charge $c=2$ and emergent SU(3) symmetry. Within numerical accuracy, it remains unclear whether the direct transition extends to finite $D$ (dashed blue lines).
     }
    \label{fig:Phase diagram}
\end{figure}

Our main results are summarized in in the remarkably rich  phase diagram in Fig.~\ref{fig:Phase diagram}, that comprises five gapped phases: the topological Haldane phase, the large-$D$ phase, an Ising-AFM, and trimerized and dimerized phases.
The Haldane and large-$D$ phases lack local order and are distinguished by their topological properties, with the Gaussian transition separating them \cite{ercolessi,PhysRevB.67.104401,PhysRevB.84.220402,s53y-qmr4}. By contrast, The Ising-AFM, trimerized, and dimerized phases spontaneously break translation symmetry and can be identified by the corresponding local order parameters (see End Matter for examples). Furthermore, we discovered two floating phases that are the main focus of this Letter.

\begin{figure}[h!]
    \centering
    \includegraphics[width=1\linewidth]{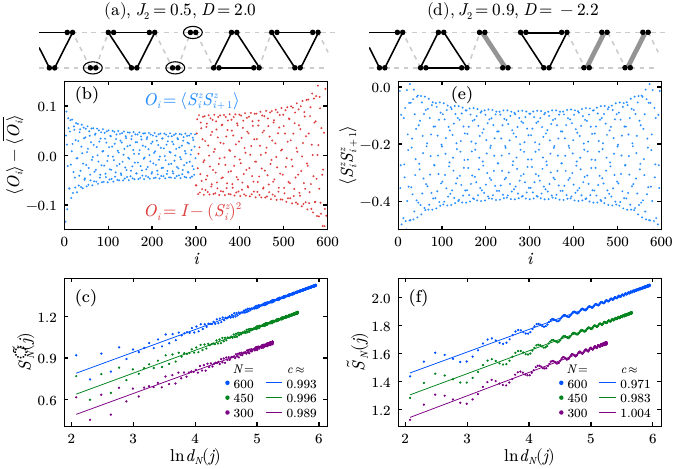}
    \caption{ Visualization and numerical evidences of emergence of (a)-(c) Floating-1 and (d)-(f) Floating-2 phases. (a),(d) Valence bond singlet (VBS) sketches of proliferation of (a) $S^z_i=0$ in the Floating-1 phase for $D>0$ and (b) dimers in the Floating-2 phase for $D<0$ over a trimerized state. (c),(f) Friedel oscillations profile of nearest-neighbor correlations (blue) inside  (c) the Floating-1 phase with $q\approx0.458\pi\in(0 , 2\pi/3)$; and (f)  the Floating-2 phase with $q\approx0.883\pi\in(2\pi/3,\pi)$. In (b) we also present local density of zeros $O_i$ (red), the wave-vector $q$ is the same for both local observables. (b),(e) Finite-size scaling of the reduced entanglement entropy with the conformal distance inside (b) Floating-1 and (e) Floating-2 phases. In both cases the  central charge extracted with Calabrese-Cardy formula (\ref{eq: Calabrese-Cardy}) is in excellent agreement with $c=1$ of Luttinger liquid. Data for $N=450$ and $N=600$ are shifted vertically for visual clarity.    
      }
    \label{fig:floating phases}
\end{figure}

When $D>0$, the anisotropy suppresses  $S^z_i=\pm1$ degrees of freedom and ultimately  stabilizes the large-D phase. As $D$ increases, effectively acting as a chemical potential, $S_i^z=0$ states proliferate between trimers, with their density saturating upon entering the large-$D$ phase. This mechanism is illustrated with valence-bond-singlets (VBS) in Fig.~\ref{fig:floating phases}(a). We corroborate this scenario by analyzing Friedel oscillations in local observables, $\langle S_i^z S_{i+1}^z\rangle$ and $(S_i^z)^2$, in Fig.~\ref{fig:floating phases}(b), which clearly reflect the presence of incommensurability. The resulting Floating-1 phase is a critical Luttinger liquid with central charge $c=1$.
We confirm this in Fig.~\ref{fig:floating phases}(c) by fitting the reduced entanglement entropy $\tilde{S}_N(j) = S_N(j) - \zeta \langle S^z_i S^z_{i+1} \rangle$, 
obtained by removing Friedel oscillations with some non-universal $\zeta$,  to the Calabrese–Cardy formula \cite{calabreseEntanglementEntropyQuantum2004}
\begin{equation}\label{eq: Calabrese-Cardy}
    S_N(j) = \frac{c}{6} \ln d_N(j) + s_1 + \ln(g).
\end{equation}
Here $d_N(j) =\frac{2N}{\pi}\sin\left( \frac{\pi j}{N}\right)$ is the conformal distance, $\ln g$ the boundary entropy and $s_1$ a non-universal constant.
 The extracted central charge $c$ agrees within within $1\%$ with field theory expectation. The Floating-1 phase is stable for LL exponents ranging between two critical values: $K^c=2/p^2=2/9$ below which the perturbation of period $p=3$ phase becomes relevant \cite{ThierryGiamargi}; and $K^c=1$, the free-fermion point beyond which the system  enters the trivial large-D phase (see End Matter for numerical results for both transitions).

Similarly,  for $D<0$ the Floating-2 phase appears as dimers favored by strong and negative $D$ proliferates through the trimers, as sketched in Fig.\ref{fig:floating phases}(d), until their maximum saturation in the dimerized phase.
 We detect this in the Friedel oscillations profile featuring standing density waves as demonstrated in  Fig.\ref{fig:floating phases}(e) with the dominant wave-vector now being in the range $2\pi/3<q<\pi$. Numerically extracted central charge is in reasonable agreement with LL $c=1$ as exemplified by Fig.\ref{fig:floating phases}(f). As indicated above, the floating phase is destroyed by trimerization once the LL exponent falls below $K^c=2/9$ \cite{ThierryGiamargi}. Based on the pronounced jumps in various quantities (see examples in the End Matter) we identified the transition into the dimerized phase to be first order.

\begin{figure}[!t]
    \centering
    \includegraphics{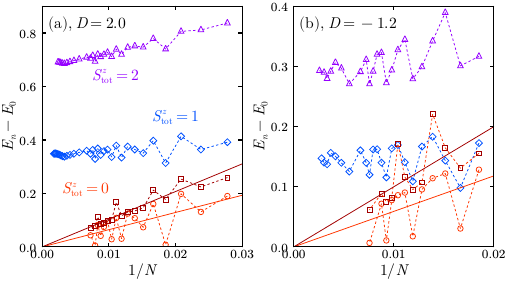}
    \caption{
        Examples of the finite-size scaling of the two low-lying singlet excitations (red), and singlet-triplet (blue) and singlet-quintuplet (purple) energy gaps inside (a) Floating-1 and (b) Floating-2 phases.
        Data is shown for $J_2=0.9$. Lines are linear fits through origin.
    }
    \label{fig:energy gap}
\end{figure}

Remarkably, both floating phases discovered here are non-magnetic: the system is critical only within the singlet sector, while magnetic excitations remain gapped across these phases. We confirm this numerically in Fig.\ref{fig:energy gap} by a direct comparison of finite-size scaling of low-lying excitations in different magnetization sectors at representative points in both floating phases. 
Although incommensurate oscillations induce strong fluctuations in the energy gaps, excitations in the $S^z_\mathrm{tot}=0$ sector scale to zero in the thermodynamic limit, while magnetic gaps remain finite (see End Matter for complementary results on spin-spin and dimer-dimer correlations). This behavior is consistent with the picture of proliferattion of magnetically neutral objects — $S_i^z=0$ states in the floating-1 phase and dimers in the floating-2 phase — between three-site singlets (trimers).
Equivalently, the domain walls between trimerized and large-$D$ or dimerized phases do not carry magnetic spinons. As a result, the condensation of these domain walls at quantum criticality does not lead to the closing of the magnetic gap. 

In contrast, domain walls between the Haldane phase and the large-$D$, trimerized, or dimerized phases carry a spinon emergent at the boundary of the topologically non-trivial Haldane domain \cite{chepigaDimerizationTransitionsSpin12016a}. This picture is fully consistent with the previously established Gaussian transition between the Haldane and large-$D$ phases \cite{ercolessi,PhysRevB.67.104401,PhysRevB.84.220402,s53y-qmr4}. However, the same argument is applicable to both floating phases: regardless the specific composition of floating domains as long as they remain non-magnetic there will be a localized spinon at their interface with the Haldane domain. In other words, the magnetic gap closes only at the transition and re-opens on another side of it, while the singlet gap once closed at the transition remains so throughout the floating phases. 

\begin{figure}[!t]
    \centering
    \includegraphics{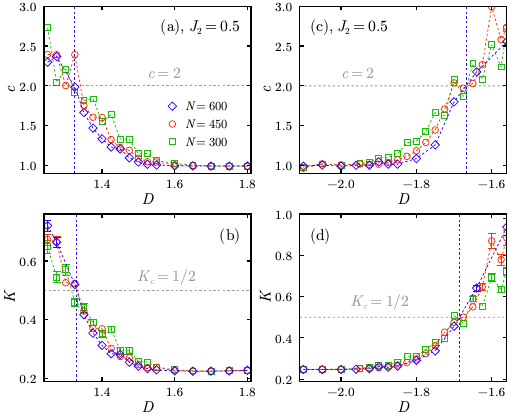}
    \caption{
        Numerical data suggesting a single topological transition with central charge $c=2$ between the Haldane phase and (a)-(b) the Floating-1 and (c)-(d) Floating-2 phases. All results are taken for $J_2=0.5$ and three different chain lengths $N$ (different colored symbols). Dashed blue lines mark estimated location of the transition. The transition out of the floating phase associated with LL parameter $K$ crossing the critical value $K^c=1/2$ corresponds to the point where the central charge reaches the value $c=2$. In the Haldane phase $K$ and $c$ should be treated as a effective  finite-size estimate of the critical exponent and central charge correspondingly. 
    }
    \label{fig:c=2}
\end{figure}

For $|D|>0$ the Haldane phase reduces to a topologically nontrivial $\mathbb{Z}_2$ phase\footnote{The four quasi-degenerate states of the Haldane phase—the singlet and the Kennedy triplet—split into two pairs, one of which is energetically favored depending on the sign of $D$.}. This allows us to identify a critical Luttinger parameter $K^c=2/p^2=1/2$, above which $\mathbb{Z}_2$ perturbations become relevant and destabilize the floating phase. At the same time, a second transition -- of Gaussian (topological) type -- must terminate the Haldane phase controlling the condensation of magnetic domain walls. Within our numerical accuracy, these two transitions appear to coincide, forming a composite critical liquid with central charge $c=2$.  In Fig.\ref{fig:c=2} we show the LL parameter $K$ and central charge $c$ across the transitions between the Haldane and both floating phases. The points where $K$ reaches its critical value $K^c=1/2$ and where the central charge crosses $c=2$ are in remarkable agreement with each other, supporting the scenario of a single transition\footnote{Within the Haldane phase, neither $K$ nor $c$ are strictly defined, and the reported values should be viewed as finite-size estimates. Notable, central charge extracted with open boundary conditions is not maximal at the transition but varies monotonously in its vicinity.}.

In summary, we have shown that single-ion anisotropy destroys the trimerized order via  proliferation of non-magnetic domain walls, giving rise to two floating phases confined to the zero-magnetization sector, while magnetic excitations remain gapped. This mechanism alone, however, is insufficient to destabilize the symmetry-protected topological Haldane phase. Instead, the transition is governed by a composite critical line with central charge $c=2$, which can be understood as the coexistence of two components: a Gaussian transition associated with gapless magnetic excitations and a non-magnetic Luttinger liquid originating from the floating phases. We hope that our results will stimulate further field-theoretical analysis to clarify whether these components are decoupled or fuse into a higher-symmetry critical theory.

From the global perspective of the phase diagram, the emergence of a magnetic LL at the boundary of the Floating-1 phase is a natural extension of the 
Gaussian transition between the Haldane and large-$D$ phases. But, if continuous, there is no other path for this transition except going through the Haldane-trimerized transition, naturally continuing along the boundary of the Floating-2 phase. Although the first-order transition between the dimerized and the Floating-2 phases  is responsible for a kink in Gaussian critical line, it continues further  as the eight-vertex transition between the Ising-AFM and dimerized phases \cite{s53y-qmr4}.

\begin{figure}[!t]
    \centering
    \includegraphics[width=1\columnwidth]{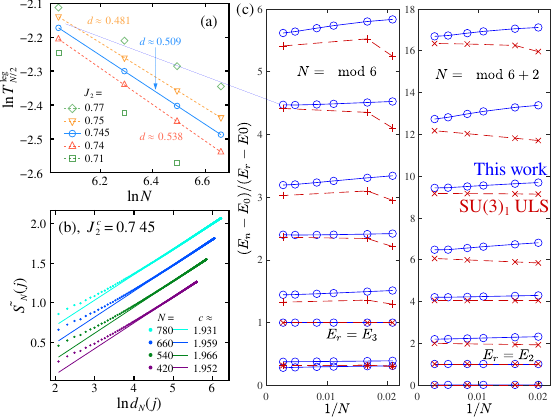}
    \caption{Numerical results for Haldane-trimerized transition. (a) Finite-size scaling of the order parameter (trimerization). We associate the critcal point with the separatrix at $J_2\approx0.745$. The slope corresponds to the effective scaling dimension $d \approx 0.509$. 
   (b) Scaling of the reduced entanglement entropy with conformal distance $d_n(j)$ (dots) fitted to the Calabrese-Cardy formula (\ref{eq: Calabrese-Cardy}). The central charge extracted numerically agree within 4$\%$ with $c=2$. (c) Conformal towers of states at the  Haldane-trimerized transition (blue) and at the SU(3)-symmetric  Uimin-Lai-Sutherland (ULS) point (red). $E_n$ are the lowest energy levels with $S^z_\mathrm{tot}=n$. The towers with $N=\mod 6$ and  $N=\mod 6+2$ are normalized with respect to reference levels $r=3$ and $r=2$ correspondingly.
    }
    \label{fig:WZW point}
\end{figure}

The present findings support the earlier prediction of a $c=2$ self-dual sine-Gordon  transition between the Haldane and trimerized phases \cite{lecheminantSUNSelfdualSine2006}. In that work, however, the transition was argued to renormalize to first order, a conclusion later challenged by numerical results suggesting critical behavior \cite{mashikoCriticalPhenomenaSU32023}. Our results further support the continuous scenario.
More specifically, along the $D=0$ line we locate the transition via finite-size scaling of the trimerization $T_i^{\mathrm{leg}}$, defined as the amplitude of next-nearest-neighbor spin correlations $\langle \mathbf{S}_i \cdot \mathbf{S}_{i+2} \rangle$ over three consecutive sites. To minimize boundary effects, we evaluate this quantity at the center of the chain ($i=N/2$). The critical point appears as a separatrix in a log–log plot in Fig.~\ref{fig:WZW point}(a), with a slope corresponding to an effective scaling dimension $d$. The central charge extracted at this point in Fig.~\ref{fig:WZW point}(b) agree with $c=2$ within $4\%$. 

Quite remarkably, the structure of the excitation spectrum at this critical point closely resembles that at the exactly solvable SU(3)-symmetric ULS point at $\theta=\pi/4$ and $J_2=0$. In Fig.~\ref{fig:WZW point}(c) we directly compare the two spectra.
One can notice strong finite-size effects, caused by the presence of marginal perturbations, responsible, in particular, for deviation of measured $c$ and $d$  from the universal values for WZW SU(3)$_1$ - $c=2$ and $d=2/3$.  
But beyond that, the two spectra show a spectacular agreement upon approaching the thermodynamic limit. This suggests an emergent SU(3) symmetry at the Haldane-trimerized transition. This is, in fact, fully consistent with commensurate  wave-vector $q=2\pi/3$ identified numerically all along $D=0$ line. Furthermore, the WZW SU(3)$_1$ critical theory is known to be composed of two boson fields\cite{difrancescoConformalFieldTheory1997} which is fully n line with the $c=2$ two-liquid transition discovered above. The extent of this SU(3) critical line (dashed line in Fig.\ref{fig:Phase diagram}) is left for future investigation.

{\bf Acknowledgments.}
NC is indebted to F.Essler for a stimulating discussion on $c=2$ line. We acknowledge useful discussions with Z. Jouini, F. Mila. BLR acknowledges useful discussion with D. Schuricht.  This work was supported by Delft Technology Fellowship, by the Royal Society (grant number URFR1251326) and  by the European Union through the ERC grant (TRANGINEER, 101220181). The views and opinions expressed are those of the authors only and do not necessarily reflect those of the European Union or the European Research Council Executive Agency; neither the European Union nor the granting authority can be held responsible for them. Numerical simulations were performed at the DelftBlue HPC and at
the Dutch national e-infrastructure with the support of the SURF Cooperative.


\clearpage

\section{End Matter}

{\bf Details of the algorithm.} We employ the two-site DMRG \cite{whiteDensityMatrixFormulation1992a, whiteDensitymatrixAlgorithmsQuantum1993a} algorithm written in terms of matrix product states (MPS) \cite{schollwoeckDensitymatrixRenormalizationGroup2011a}. We study chains of $N=3k$ sites, with $k$ denoting the number of trimers (except for the spectrum in Fig.\ref{fig:WZW point}(c)).
Unless stated otherwise, we leave the edges of the chain open, keep a maximum up to generically $2000$ and eventually $4000$ states, discard Schmidt values smaller than $10^{-8}$, and let the algorithm run till the absolute error in the energy of subsequent sweeps is smaller than $10^{-7}$.
Excitation energies within the $S^z_\mathrm{tot}=0$ sector were obtained by targeting multiple states as described in Ref.\cite{chepigaExcitationSpectrumDensity2017}.

{\bf Additional details on gapped phases:}

\begin{figure}[!h]
    \centering
    \includegraphics[width=0.48\textwidth]{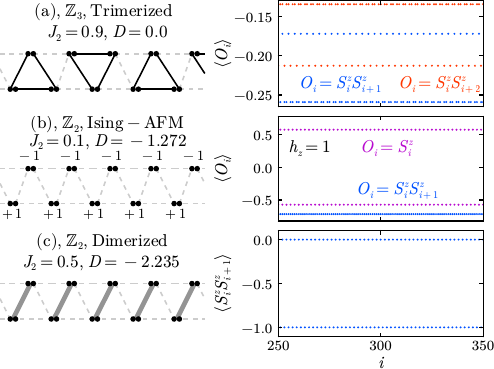}
    \vspace{-0.7cm}
\caption{VBS sketches and local observables extracted in the three ordered phases: (a) Trimerized phase with period-3 alternating NN $\langle S_i^z S_{i+1}^z \rangle$ and NNN $\langle S_i^z S_{i+2}^z \rangle$;   (b) Alternating local magnetization $S^z_i$ and uniform  NN correlations in Ising-AFM phase; and 
        (c) Alternating NN correlations in the  dimerized phase.
          }
    \label{fig:Sketches}
\end{figure}

{\it Trimerized phase} spontaneously breaks the $\mathbb{Z}_3$ translation symmetry. In Fig.\ref{fig:Sketches}(a) we present both a VBS sketch and a pattern of numerically computed local correlations on the legs and rungs that support this interpretation.

{\it Ising-AFM phase} spontaneously breaks $\mathbb{Z}_2$ symmetry resulting in the alternating magnetization  while rung correlations are uniform  as shown in Fig.\ref{fig:Sketches}(b).

{\it Dimerized phase} spontaneously breaks $\mathbb{Z}_2$ symmetry on bonds, detected in Fig.\ref{fig:Sketches}(c) through alternating NN correlations. 

\vspace{-0.3cm}
{\it Haldane phase} is a topologically non-trivial uniform phase.  The phase has incommensurate short-range order except a narrow region near the Gaussian transition where it is commensurate with $q=\pi$. Within numerical accuracy we see $q=2\pi/3$ all along $D=0$ line.

{\it Large-D phase} is commensurate almost everywhere with $q=0$ except for a narrow region near the Floating-1 phase with incommensurate short-range correlations.

\begin{figure}[!h]
    \centering
    \includegraphics[width=0.48\textwidth]{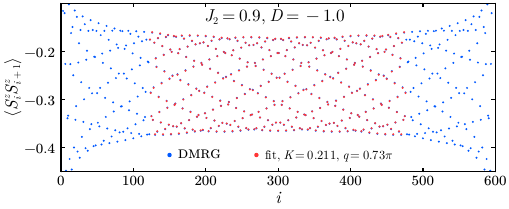}
        \vspace{-0.7cm}
    \caption{
       Example of fitting nearest-neighbor correlations $\langle S^z_i S^z_{i+1} \rangle$ against Eq.\eqref{eq:Friedel oscillations} in the floating phase.
    }
    \label{fig:extractKq}
\end{figure}

{\bf Extracting LL exponent $K$.}
Open boundary conditions act as impurities and induce Friedel oscillations in local observables like NN correlations $\langle S^z_i S^z_{i+1} \rangle$. According to boundary conformal field theory\cite{chepigaLifshitzPointCommensurate2021}:
\begin{equation}\label{eq:Friedel oscillations}
    \langle S^z_i S^z_{i+1} \rangle \propto \frac{\cos(q i + \phi_1)}{\left[ (N/\pi) \sin(\pi i / N) \right]^K},
\end{equation}
where $K$ and $q$ are the LL parameter and incommensurate wave vector respectively, $\phi_1$ is a non-universal phase. To reduce finite-size effects, we discard edge sites and fit only the central part of the chain as shown in Fig.\ref{fig:extractKq}, typically removing 10--25\% of sites on each end, and taking $K$ as a mean value with respect to discarded intervals.

\begin{figure}[!h]
    \centering
    \includegraphics[width=0.48\textwidth]{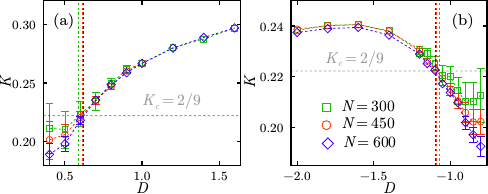}
        \vspace{-0.7cm}
    \caption{ Luttinger liquid exponent $K$ across  Kosterlitz-Thouless transition between gapped trimerized phase and (a) Floating-1 and (b) Floating-2 phases along $J_2=0.9$. The transitions takes place when LL exponent reaches critical value $K^c=2/9$ (vertical dotted lines).         
    }
    \label{fig:floatingtrimerized}
\end{figure}

{\bf Kosterlitz-Thouless transitions into trimerized phase.} Both Floating-1 and 2 phases are destroyed when trimer perturbation become relevant. This happens when the LL exponent is below $K^c=2/9$\cite{ThierryGiamargi,kosterlitzOrderingMetastabilityPhase1973a}. In Fig.\ref{fig:floatingtrimerized} we show an example for $J_2=0.9$, where the trimerized phase spans between $-1.1 \lesssim D \lesssim 0.6$. Inside the trimerized phase $K$ is not defined and should be understood an effective finite-size estimate, indeed showing a strong finite-size dependence in the gapped phases.

\begin{figure}[!h]
    \centering
    \includegraphics[width=0.48\textwidth]{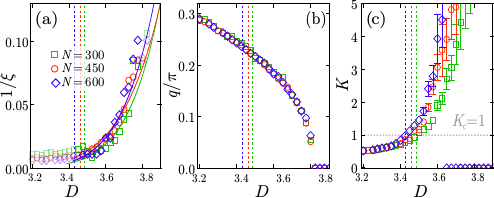}
        \vspace{-0.7cm}
    \caption{(a) Inverse of the correlation length, (b) incommensurate wave-vector $q$, and LL exponent $K$ across the Kosterlitz-Thouless transition between the Floating-1 and large-D phases along  $J_2=0.9$. We only show error bars that exceed the sizes of the symbols. Vertical lines indicate the finite-size location of the transition estimated from $K=K^c=1$ in (c).
    }
    \label{fig:larged}
\end{figure}

{\bf Kosterlitz-Thouless transitions into Large-D phase.} The density of $S^z_i=0$ state saturates in the trivial phase and the transition takes place when LL parameter $K$ reaches the critical value of free-fermion $K^c=1$ (see Fig.\ref{fig:larged}(c)). The transition is concluded to be of the Kosterlitz-Thouless\cite{kosterlitzOrderingMetastabilityPhase1973a} type (as opposed to Pokrovsky-Talapov\cite{pokrovskyGroundStateSpectrum1979a}) based on the zero slope in the inverse of the correlation length (see Fig.\ref{fig:larged}(a)) consistent with a stretch-exponential divergence of $\xi$ and  a  wave-vector $q$ being incommensurate on both sides of the transition (see Fig.\ref{fig:larged}(b)).

\begin{figure}[!h]
    \centering
    \includegraphics{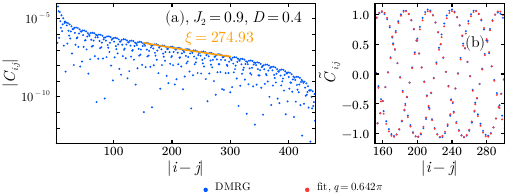}
        \caption{
        Example of fitting the dimer-dimer correlation function to the Ornstein-Zernicke form \eqref{eq:oz} in two steps:  
        (a) First, the correlation length $\xi$ is extracted from the main slope of the logarithm of $| C_{i,j} |$.
        (b) Then we fit the reduced correlations $\tilde{C}_{i,j}=C_{i,j}\cdot Ae^{|i-j|/\xi}\sqrt{|i-j|}$ and fit it with $\cos (q+\phi_0)$.
    }
    \label{fig:extractxiandq}
\end{figure}

{\bf Correlations in the Floating-1 phase.} In addition to the gap scaling in Fig.\ref{fig:energy gap}(main text) the non-magnetic nature of the floating phases is reflected in exponentially decaying spin-spin correlations. We define
dimer-dimer
$    C_\mathrm{rung}(i,j) = \langle S^z_{i} S^z_{i+1} S^z_{j} S^z_{j+1} \rangle - 
            \langle S^z_{i} S^z_{i+1} \rangle \langle S^z_{j} S^z_{j+1} \rangle$
and spin-spin
$
    C_{S^+-S^-}(i,j) = \langle S^+_{i}S^-_{j} \rangle
$ correlations and fit them against the Ornstein-Zernicke form\cite{ornsteinAccidentalDeviationsDensity1994}:
\begin{equation}\label{eq:oz}
    C^{\mathrm{OZ}}(i,j) \propto \frac{e^{-|i-j|/\xi}}{\sqrt{|i-j|}} \cos(q|i-j|+\phi_0),
\end{equation}
where $q$ is the wavevector, and $\phi_0$ is a non-universal phase. Example of the fit is shown in Fig.\ref{fig:extractxiandq}.

\begin{figure}[!h]
    \includegraphics[width=0.48\textwidth]{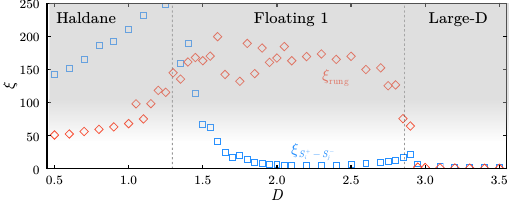}
        \vspace{-0.2cm}
    \caption{ Correlation length extracted from the decay of dimer-dimer (red) and spin-spin (blue) correlations along $J_2=0.5$. Spin-spin correlation length remains finite across the non-magnetic Floating-1 phase but diverges at the $c=2$ transition to the Haldane phase.  Shaded area indicate that the results should be treated with caution as correlation lengths are comparable to the size of the system ($N=450$).     Dashed vertical lines denote estimates of the transitions.
    }
    \label{fig:corlen}
\end{figure}

Our results are summarized in Fig.\ref{fig:corlen}  where we contrast the spin-spin and dimer-dimer correlations. Spin-spin correlation length remains finite in the Floating-1 phase and dimerges only upon approaching the transition to the Haldane phase, while dimer-dimer correlation length is comparable to the system size it was extracted from consistent with critical regime. The resutls presented in Fig.\ref{fig:corlen} further support {(\it i)} non-magnetic nature of the Floating-1 phase; and {\it (ii)} an additional magnetic criticality due to divergent spin-spin correlation length at the transition to the Haldane phase.

\begin{figure}[!b]
    \centering
    \includegraphics[width=0.48\textwidth]{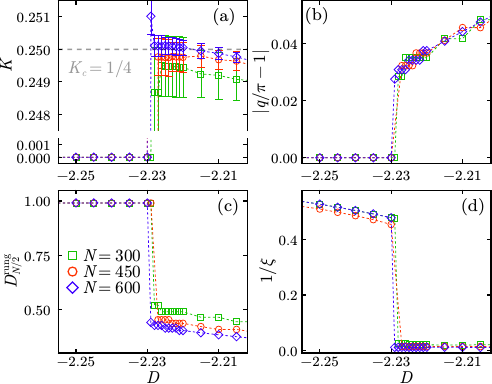}
    \vspace{-0.2cm}
    \caption{Finite jump in (a) Luttinger liquid parameter, (b) incommensurate wave-vector, (c) alternation in nearest-neighbor correlations $D^\mathrm{rung}_{N/2}=|S^z_{N/2-1}S^z_{N/2}-S^z_{N/2}S^z_{N/2+1}|$, and (d) inverse of the correlation length at $J_2=0.5$ supporting first order transition between the Floating-2 and the dimerized phase. 
    }
    \label{fig:first order}
\end{figure}

{\bf First order transition} between Floating-2 and  dimerized phases is seen through pronounced jumps in the order parameter, correlation length and incommensurability as shown in Fig.\ref{fig:first order}. Interestingly enough, the LL exponent $K$, that abruptly drops to zero in the dimerized phase, reaches the value $K=1/4$ of the Pokrovsky-Talapov transition\cite{pokrovskyGroundStateSpectrum1979a} into the period-2 phase.

\clearpage

\bibliography{references}

\clearpage

\end{document}